\documentclass[preprint,aps,showpacs]{revtex4}
\usepackage{mathrsfs}
\usepackage{amsmath}
\usepackage{amssymb}
\usepackage{epsfig}
\usepackage{graphicx}
\usepackage{booktabs}
\usepackage{array}
\usepackage{multirow}
\begin{document}
\renewcommand{\arraystretch}{0.5}
\newcommand{\beq}{\begin{eqnarray}}
\newcommand{\eeq}{\end{eqnarray}}
\newcommand{\non}{\nonumber\\ }

\newcommand{\acp}{ {\cal A}_{CP} }
\newcommand{\psl}{ p \hspace{-1.8truemm}/ }
\newcommand{\nsl}{ n \hspace{-2.2truemm}/ }
\newcommand{\vsl}{ v \hspace{-2.2truemm}/ }
\newcommand{\epsl}{\epsilon \hspace{-1.8truemm}/\,  }

\def \cpl{ Chin. Phys. Lett.  }
\def \ctp{ Commun. Theor. Phys.  }
\def \epjc{ Eur. Phys. J. C }
\def \jpg{  J. Phys. G }
\def \npb{  Nucl. Phys. B }
\def \plb{  Phys. Lett. B }
\def \prd{  Phys. Rev. D }
\def \prl{  Phys. Rev. Lett.  }
\def \zpc{  Z. Phys. C }
\def \jhep{ J. High Energy Phys.  }

\title{The $B_c\rightarrow D^{(*)}T$ decays in perturbative QCD approach}
\author{Zhi-Tian Zou$^a$,
Xin Yu$^a$ and Cai-Dian
L\"u $^{(a,b)}$\footnote{lucd@ihep.ac.cn} } 
\affiliation{ a. Institute  of  High  Energy  Physics  and  Theoretical  Physics Center for Science Facilities,
Chinese Academy of Sciences, Beijing 100049, People's Republic of China\\
b. Kavli Institute for Theoretical Physics China, CAS, Beijing 100190, China} 

\date{\today}
\begin{abstract}
In this work, we investigate those $B_{c}\rightarrow
D^{(*)}T$ decays in perturbative QCD
approach, based on $k_T$ factorization, where T denotes
a light tensor meson. For all decays considered in this work, there are no contributions from factorizable emission diagrams because the emitted meson is the tensor meson. We find that the annihilation amplitudes are dominant in these decays due to the large Cabibbo-Kobayashi-Maskawa elements, which are only calculable in the pQCD approach. The numerical
results show that the predictions for the branching ratios of
most decays are in the order of $10^{-6}$ or even bigger, which can be observed in the ongoing experiments. We also predict large
percentage of transverse polarizations in   those W annihilation diagram dominant $B_{c} \rightarrow
D^{*}T$ decay channels.
\end{abstract}



\maketitle

\section{Introduction}
After the first observation  was reported in 1998 by the CDF collaboration \cite{cdf}, which was confirmed until 2008 by CDF and D0 collaboration \cite{cdfd} at Tevatron in excess of 5$\sigma$ significance, the study of $B_{c}$ meson is becoming one of the currently interesting topics, especially since the Large Hadron Collider (LHC) experiment ran normally. From the point of structure, the $B_{c}$ meson is a ground state of two heavy quarks' system, with a $c$ quark and a $\bar{b}$ quark, which is very different from the symmetric heavy quarkonium ($\bar{c}c,\bar{b}b$) states, due to the flavor $B=-C=\pm1$ carried by $B_{c}$ meson. Since the $B_{c}$ meson carries explicit flavor, it can not annihilate via strong interaction or electromagnetic interaction like the mesons consisting of $\bar{c}c$ or $\bar{b}b$. It can only decay via weak interaction. Thus it provides us an ideal platform to understand the weak interaction of heavy quark flavor  \cite{haochu1,haochu2}. Unlike the heavy-light $B_{q}$ meson (q= u, d, s), both the $\bar{b}$ and $c$ can decay with the other as spectator, or they annihilate into pairs of leptons or light mesons. If  more data become available, the $B_{c}$ physics must be a good place    to study the perturbative and nonperturbative QCD dynamics, final state interactions, even the new physics beyond the standard model \cite{haochu1,haochu2}. In recent years, many theoretical studies on the production and decays of $B_{c}$ meson have been done  based on Operator Production Expansion \cite{ope1,ope2}, nonrelativistic QCD (NRQCD) and perturbative methods \cite{nr1,nr2,p1,p2,p3}, QCD sum rules \cite{sr1,sr2}, SU(3) flavor symmetry \cite{su}, Isgur-Scora-Grinstein-Wise (ISGW) quark model \cite{isgw1,isgw2,isgw3}, QCD factorization approach \cite{qcdf1,qcdf2}, and the perturbative QCD (PQCD) approach \cite{pqcd1,pqcd2,pqcd3,pqcd4,pqcd5,pqcd6,pqcd7}.

The  $B$ meson decays involving a tensor meson have been studied in refs.\cite{zheng1,zheng2,wwprd83014008,prd491645,prd555581,prd59077504,epjc22683,epjc22695,prd67014002,jpg36095004,
arxiv1010.3077,prd67014011,prd85051301,zou}.  In refs.\cite{isgw2,isgw3}, the authors have studied some analogous $B_{c}$ decays involving a tensor meson in final states, but only with the tensor meson as the recoiled meson. In this work, we focus on the $B_{c}\rightarrow D^{(*)}T$ decays, where $T$ denotes a light tensor meson with $J^{P}=2^+$, which is emitted from vacuum. We know that factorizable amplitude proportional to matrix element $<T\mid j^{\mu}\mid 0>$ ,where $j^{\mu}$ is the $(V\pm A)$ or $(S\pm P)$ current, does not contribute because this matrix element vanishes from lorentz covariance considerations \cite{zheng1,zheng2,epjc22683,epjc22695}, so these $B_{c}\rightarrow D^{(*)}T$ decays are prohibited in naive factorization. To our knowledge, these decays are never considered in the theoretical papers due to this difficulty of factorization.  In order to give the predictions to these decay channels, it is necessary to go beyond the naive factorization to calculate the nonfactorizable  and annihilation diagrams. What is more, the annihilation amplitudes will be dominant in considered $B_{c}\rightarrow D^{(*)}T$ decays because they depend upon the large Cabibbo-Kobayashi-Maskawa (CKM) elements $V_{cb}$ and $V_{cs(d)}$. It is worth of mentioning that the PQCD approach  is almost the only method can do the quantitative calculations of the annihilation type diagrams \cite{annihilation1,annihilation2}. The PQCD approach have successfully predicted the pure annihilation type decays $B_{s}\rightarrow \pi^{+}\pi^{-}$ \cite{prd70034009,prd76074018} and $B^0\rightarrow D_{s}^{-}K^{+}$ \cite{epjc28305,prd78014018}, which have been confirmed by experiments later \cite{10498,jpg37075021}. So, for these annihilation dominant decays, the calculation in PQCD approach is reliable.

In this paper, we shall study these $B_{c}\rightarrow D^{(*)}T$ decays in the PQCD approach, which is based on the $k_{T}$ factorization \cite{plb504,prd63074009,pnp5185}.
In this approach, we keep the transverse momentum of quarks, and as a result, the end-point singularity in collinear factorization can be avoided. On the other hand, the double logarithms will appear in QCD correction due to the additional energy scale introduced by the transverse momentum. Using the renormalization group equation, the double logarithms can be resumed, which results in the Sudakov form factor. This factor effectively suppresses the end point contribution of the distribution amplitude of mesons in the small transverse momentum region, which makes the calculation in PQCD appraoch reliable and consistent.

In these decays, there is one more intermediate energy scale,
the D meson mass. As a result, another expansion series of
$m_{D}/m_{B_c}$ will appear. The factorization is only approved at the
leading of $m_{D}/m_{B_c}$ expansion \cite{fd1,fd2}, which has also been proved by soft collinear effective theory \cite{scet}. Therefore, we
will take only the leading order contribution in account, unless
explicitly mentioned.

This paper is organized as follows. In Sec.II, we present the
formalism and wave functions of the considered decays. Then
we perform the perturbative calculations for considered decay
channels with the PQCD approach in Sec.III. The numerical results and
phenomenological analysis are given in Sec.IV. Finally, Sec.V
contains a short summary.

\section{FORMALISM AND WAVE FUNCTION}\label{sec:function}

In order to give the predictions for these considered $B_{c}\rightarrow D^{(*)}T$ decays, the key step is to calculate the transition matrix elements:
\begin{eqnarray}
\mathcal{M}\propto \langle D^{(*)}T|\mathcal{H}_{eff}|B_{c}\rangle
\end{eqnarray}
where the weak effective Hamiltonian $\mathcal{H}_{eff}$ can be written as \cite{rmp681125}
\begin{eqnarray}
\mathcal{H}_{eff}=&&\frac{G_{F}}{\sqrt{2}}\left\{\sum_{q=u,c} V_{qb}^{*}V_{qX}\left[C_{1}(\mu)O_{1}^{q}(\mu)+C_{2}(\mu)O_{2}^{q}(\mu)\right]\right.\nonumber\\
&&\left. -V_{tb}^{*}V_{tX}\left[\sum_{i=3}^{10}C_{i}(\mu)O_{i}(\mu)\right]\right\},
\end{eqnarray}
with $V_{qb(X)}$ and $V_{tb(X)}$  ($X=d,s$) the CKM matrix elements. $O_{j}\,(j=1,...,10)$ are the local four-quark operators:

current-current (tree) operators
\begin{eqnarray}
O_{1}^{q}=(\bar{b}_{\alpha}q_{\beta})_{V-A}(\bar{q}_{\beta}X_{\alpha})_{V-A},\;\;\;O_{2}^{q}=(\bar{b}_{\alpha}q_{\alpha})
_{V-A}(\bar{q}_{\beta}X_{\beta})_{V-A},
\end{eqnarray}

QCD penguin operators
\begin{eqnarray}
&&O_{3}=(\bar{b}_{\alpha}X_{\alpha})_{V-A}\sum_{q^{\prime}}(\bar{q}^{\prime}_{\beta}q^{\prime}_{\beta})_{V-A},\;\;\;
O_{4}=(\bar{b}_{\alpha}X_{\beta})_{V-A}\sum_{q^{\prime}}(\bar{q}^{\prime}_{\beta}q^{\prime}_{\alpha})_{V-A},\\
&&O_{5}=(\bar{b}_{\alpha}X_{\alpha})_{V-A}\sum_{q^{\prime}}(\bar{q}^{\prime}_{\beta}q^{\prime}_{\beta})_{V+A},\;\;\;
O_{6}=(\bar{b}_{\alpha}X_{\beta})_{V-A}\sum_{q^{\prime}}(\bar{q}^{\prime}_{\beta}q^{\prime}_{\alpha})_{V+A},
\end{eqnarray}

electro-weak penguin operators
\begin{eqnarray}
&&O_{7}=\frac{3}{2}(\bar{b}_{\alpha}X_{\alpha})_{V-A}\sum_{q^{\prime}}e_{q^{\prime}}(\bar{q}^{\prime}_{\beta}q^{\prime}_{\beta})_{V+A},
\;\;O_{8}=\frac{3}{2}(\bar{b}_{\alpha}X_{\beta})_{V-A}\sum_{q^{\prime}}e_{q^{\prime}}(\bar{q}^{\prime}_{\beta}q^{\prime}_{\alpha})_{V+A},\\
&&O_{9}=\frac{3}{2}(\bar{b}_{\alpha}X_{\alpha})_{V-A}\sum_{q^{\prime}}e_{q^{\prime}}(\bar{q}^{\prime}_{\beta}q^{\prime}_{\beta})_{V-A},\;\;
O_{10}=\frac{3}{2}(\bar{b}_{\alpha}X_{\beta})_{V-A}\sum_{q^{\prime}}e_{q^{\prime}}(\bar{q}^{\prime}_{\beta}q^{\prime}_{\alpha})_{V-A},
\end{eqnarray}
where $\alpha$ and $\beta$ are the color indices and $q^{\prime}$ are the active quarks at the scale $m_{b}$, i. e. $q^{\prime}=(u,d,s,c,b)$. The left-handed and right-handed currents are defined as $(\bar{b}_{\alpha}q_{\beta})_{V-A}=\bar{b}_{\alpha}\gamma_{\mu}(1-\gamma_{5})q_{\beta}$ and $(\bar{q}^{\prime}_{\beta}q^{\prime}_{\alpha})_{V+A}=\bar{q}^{\prime}_{\beta}\gamma_{\mu}(1+\gamma_{5})q^{\prime}_{\alpha}$ respectively. The combinations $a_{i}$ of the Wilson coefficients are defined as \cite{prd58094009}:
\begin{eqnarray}
&&a_{1}=C_{2}+C_{1}/3,\;\;\;\;\;\;a_{2}=C_{1}+C_{2}/3,\nonumber\\
&&a_{i}=C_{i}+C_{i+1}/3,\,i=3,5,7,9,\;\;\;a_{j}=C_{j}+C_{j-1}/3, \,j=4,6,8,10.
\end{eqnarray}

In hadronic $B$ decays, there are several typical scales, and expansions with respect to the ratios of the scales are ususlly carried out. The electroweak physics higher than W boson mass can be calculated perturbatively. The physics between b quark mass scale and W boson mass scale can be included in the above Wilson coefficients $C_i(\mu)$ of the effective four-quark operators, which is obtained by using the renormalization group equation. The physics between $M_{B}$ and the factorization scale is included in the calculated hard part in the PQCD approach. The physics below the factorization scale is nonperturbative and described by the hadronic wave functions of mesons, which is universal for all decay modes. Finally, in the PQCD approach, the decay amplitude can be factorized into the
convolution of the the Wilson coefficients $C(t)$, the hard scattering
kernel and the light-cone wave functions $\Phi_{M_{i},(B)}$ of mesons characterized by
different scales,
\begin{eqnarray}
\mathcal
{A}\;\sim\;&&\int\,dx_{1}dx_{2}dx_{3}b_{1}db_{1}b_{2}db_{2}b_{3}db_{3}\nonumber\\
&&\times
Tr\left[C(t)\Phi_{B}(x_{1},b_{1})\Phi_{M_{2}}(x_{2},b_{2})\Phi_{M_{3}}(x_{3},b_{3})H(x_{i},b_{i},t)S_{t}(x_{i})e^{-S(t)}\right],
\end{eqnarray}
where  $b_{i}$ is the conjugate variable of quark's transverse
momentum $k_{iT}$, $x_{i}$ is the momentum fractions of valence quarks and $t$ is the largest scale in  the hard part
$H(x_{i},b_{i},t)$.   The jet function
$S_{t}(x_{i})$, which is obtained by the threshold resummation, smears the end-point singularities on $x_{i}$
\cite{prd66094010}. The Sudakov form
factor  $e^{-S(t)}$ is  from the resummation of the double logarithms, which suppresses the soft dynamics effectively   i.e.
the long distance contributions in the large $b$ region
\cite{prd57443,lvepjc23275}. Thus it makes the perturbative
calculation of the hard part $H$ applicable at intermediate scale,
i.e., $m_{B}$ scale.


In the PQCD approach,   the initial and final state meson wave functions are the most important non-perturbative inputs. For $B_{c}$ meson, we only consider the contribution from the first Lorentz structure, like $B_{q}\,(q=u,d,s)$ meson,
\begin{eqnarray}
\Phi_{B_{c}}(x)=\frac{i}{\sqrt{2N_{c}}}(\makebox[-1.5pt][l]{/}P+m_{B_{c}})\gamma_{5} \phi_{B_c}(x,b).
\end{eqnarray}
For the distribution amplitude, we adopt the model \cite{pqcd1}:
\begin{eqnarray}
\phi_{B_c}(x,b)=\frac{f_{B_c}}{2\sqrt{2N_c}}\,\delta(x-m_{c}/m_{B_c})\exp\left[-\frac{1}{2}w^2b^2\right],
\end{eqnarray}
in which $\exp\left[-\frac{1}{2}w^2b^2\right]$ represents the $k_{T}$ dependence. $f_{B_c}$ and $N_c=3$ are the decay constant of $B_c$ meson and the color number respectively.

As discussion in ref.\cite{wwprd83014008}, for these $B_{c}\rightarrow D^{(*)}T$ decays, the $\pm2$ polarizations ($J^P=2^+$) do not contribute due to the angular momentum conservation argument. Because of the simplification, the wave functions for a generic tensor meson are defined
by \cite{wwprd83014008}
\begin{eqnarray}
&&\Phi_{T}^{L}\,=\,\frac{1}{\sqrt{6}}\left[m_{T}\makebox[0pt][l]{/}\epsilon_{\bullet
L}^{*}\phi_{T}(x)\,+\,\makebox[0pt][l]{/}\epsilon_{\bullet
L}^{*}\makebox[-1.5pt][l]{/}P\phi_{T}^{t}(x)+m_{T}^{2}\frac{\epsilon_{\bullet}\cdot
v}{P\cdot v}\phi_{T}^{s}(x)\right]\nonumber\\
&&\Phi_{T}^{\perp}\,=\,\frac{1}{\sqrt{6}}\left[m_{T}\makebox[0pt][l]{/}\epsilon_{\bullet
\perp}^{*}\phi_{T}^{v}(x)\,+\,\makebox[0pt][l]{/}\epsilon_{\bullet
\perp}^{*}\makebox[-1.5pt][l]{/}P\phi_{T}^{T}(x)\,+\,m_{T}i\epsilon_{\mu\nu\rho\sigma}\gamma_{5}\gamma^{\mu}\epsilon_{\bullet
\perp}^{* \nu}n^{\rho}v^{\sigma}\phi_{T}^{a}(x)\right],
\end{eqnarray}
where
$\epsilon_{\bullet}\,\equiv\,\frac{\epsilon_{\mu\nu}v^{\nu}}{P\cdot\,
v}$, and $\epsilon_{\mu\nu}$ is the polarization tensor, which can be found in refs.\cite{zheng1,zheng2,wwprd83014008}. The distribution
amplitudes can be given by \cite{wwprd83014008,zheng1,zheng2}
\begin{eqnarray}
&&\phi_{T}(x)\,=\,\frac{f_{T}}{2\sqrt{2N_{c}}}\phi_{\|}(x),\;\phi_{T}^{t}\,=\,\frac{f_{T}^{\perp}}{2\sqrt{2N_{c}}}h_{\|}^{(t)}(x),
\nonumber\\
&&\phi_{T}^{s}(x)\,=\,\frac{f_{T}^{\perp}}{4\sqrt{2N_{c}}}\frac{d}{dx}h_{\|}^{(s)}(x),\;\phi_{T}^{T}(x)\,=\,\frac{f_{T}^{\perp}}{2\sqrt{2N_{c}}}\phi_{\perp}(x),\nonumber\\
&&\phi_{T}^{v}(x)\,=\,\frac{f_{T}}{2\sqrt{2N_{c}}}g_{\perp}^{(v)}(x),\;\phi_{T}^{a}(x)\,=\,\frac{f_{T}}{8\sqrt{2N_{c}}}\frac{d}{dx}g_{\perp}^{(a)}(x).
\end{eqnarray}
The asymptotic twist-2 and twist-3 distributions are: \cite{zheng1,zheng2,wwprd83014008}
\begin{eqnarray}
&&\phi_{\|,\perp}(x)\,=\,30x(1-x)(2x-1),\nonumber\\
&&h_{\|}^{(t)}(x)\,=\,\frac{15}{2}(2x-1)(1-6x+6x^{2}),\;h_{\|}^{(s)}(x)\,=\,15x(1-x)(2x-1),\nonumber\\
&&g_{\perp}^{(a)}(x)\,=\,20x(1-x)(2x-1),\;\;g_{\perp}^{(v)}(x)\,=\,5(2x-1)^{3}.
\end{eqnarray}
These light-cone distribution amplitudes (LCDAs) of the light tensor meson are asymmetric under the interchange of momentum fractions of quark and anti-quark in the SU(3) limit because of the Bose statistics \cite{zheng1,zheng2}.

For $D^{(*)}$ meson, in the heavy quark limit, the two-parton
LCDAs can be written as refs.
\cite{pqcd1,prd67054028,liprd78014018,zouhaojpg37,liprd81034006}
\begin{eqnarray}
\langle D(p)|q_{\alpha}(z)\bar{c}_{\beta}(0)|0\rangle
\,&=&\,\frac{i}{\sqrt{2N_{c}}}\int_{0}^{1}dx\,e^{ixp\cdot
z}\left[\gamma_{5}(\makebox[-1.5pt][l]{/}P\,+\,m_{D})\phi_{D}(x,b)\right]_{\alpha\beta},\nonumber\\
\langle D^{*}(p)|q_{\alpha}(z)\bar{c}_{\beta}(0)|0\rangle
\,&=&\,-\frac{1}{\sqrt{2N_{c}}}\int_{0}^{1}dx\,e^{ixp\cdot
z}\left[\makebox[-1.5pt][l]{/}\epsilon_{L}(\makebox[-1.5pt][l]{/}P\,+\,m_{D^{*}})\phi_{D^{*}}^{L}(x,b)\right.\nonumber\\
&&\left.\,+\;\makebox[-1.5pt][l]{/}\epsilon_{T}(\makebox[-1.5pt][l]{/}P\,+\,m_{D^{*}})\phi_{D^{*}}^{T}(x,b)\right]_{\alpha\beta},
\end{eqnarray}
 For the distribution amplitude for D meson, we take the same
model as that used in Refs.
\cite{liprd78014018,zouhaojpg37,liprd81034006}.
\begin{eqnarray}
\phi_{D}(x,b)\,=\,\frac{1}{2\sqrt{2N_{c}}}\,f_{D}\,6x(1-x)\left[1+C_{D}(1-2x)\right]
\exp\left[\frac{-\omega^{2}b^{2}}{2}\right],
\end{eqnarray}
with $C_{D}=0.5\pm0.1, \omega=0.1$ GeV and $f_{D}=207$ MeV \cite{fd}
for $D (\bar{D}$) meson and $C_{D}=0.4\pm0.1, \omega=0.2$ GeV and
$f_{D_{s}}=241$ MeV \cite{fd}  for $D_{s} (\bar{D}_{s})$ meson. For
$D^{*}$ meson, we take the same model as the $D$ meson and determine the decay constant by
using the following relation based on heavy quark effective theory
(HQET) \cite{am}.
\begin{eqnarray}
f_{D_{(s)}^{*}}\,=\,\sqrt{\frac{m_{D_{(s)}}}{m_{D_{(s)}}^{*}}}\,f_{D_{(s)}}
\end{eqnarray}

\section{Perturbative calculation}\label{jiexi}

There are 6 types of diagrams contributing to the
$B_c\rightarrow D^{(*)}T$ decays, which are shown in Fig.1. The   dominant factorizable emission type diagrams in most other decay modes are not shown here, because they do not contribute  for a tensor meson emission. The second line are the factorizable and nonfactorizable annihilation type diagrams.
\begin{figure}[]
\begin{center}
\vspace{-5cm} \centerline{\epsfxsize=10 cm \epsffile{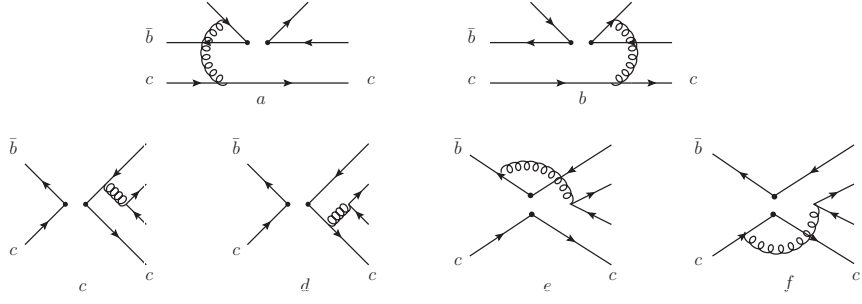}}
\vspace{-5cm} \caption{Leading order Feynman diagrams contributing to the
$B_c\,\rightarrow\,D^{(*)}T$ decays in PQCD }
 \label{fig:lodiagram}
 \end{center}
\end{figure}

After the perturbative calculation, the decay amplitudes for the non-factorizable emission diagrams in Fig.1(a) and (b) are

(i) (V-A)(V-A) operators:
\begin{eqnarray}
\mathcal{M}_{enf}^{LL}&=&\frac{32}{3}\pi C_{F}m_{B_c}^{4}\int_{0}^{1}d[x]\int_{0}^{1/\Lambda}b_1db_1b_2db_2\phi_{B_c}(x_1,b_1)\phi_{T}(x_{2})\phi_{D}(x_3,b_{1})\nonumber\\
&& \times \left\{\left[r_{D}(1-x_{3})+x_{1}+x_{2}-1\right]E_{enf}(t_{a})h_{enf}(x_{1},(1-x_{2}),x_{3},b_{1},b_{2})\right.\nonumber\\
&&\left.-\left[r_{D}(1-x_{3})+x_{1}-x_{2}+x_{3}-1\right]E_{enf}(t_b)h_{enf}(x_{1},x_{2},x_{3},b_{1},b_{2})\right\},
\end{eqnarray}

(ii) (V-A)(V+A) operators:
\begin{eqnarray}
\mathcal{M}_{enf}^{LR}&=&\frac{32}{3}\pi C_{F}r_{T}m_{B_c}^{4}\int_{0}^{1}d[x]\int_{0}^{1/\Lambda}b_1db_1b_2db_2\phi_{B_{c}}(x_{1},b_1)\phi_{D}(x_3,b_1)\nonumber\\
&&\times\left\{\left[\phi_{T}^{s}(x_2)(x_1+x_2+r_{D}(x_1+x_2+x_3-2)-1)\right.\right.\nonumber\\
&&\left.\left.+\phi_{T}^t(x_2)((x_1+x_2)(1+r_D)-r_Dx_3-1)\right]\right.\nonumber\\
&&\left.\cdot E_{enf}(t_{a})h_{enf}(x_{1},(1-x_{2}),x_{3},b_{1},b_{2})\right.\nonumber\\
&&\left.+\left[\phi_{T}^t(x_2)(x_1-x_2+r_D(x_1-x_2-x_3+1))\right.\right.\nonumber\\
&&\left.\left.-\phi_{T}^s(x_2)(x_1-x_2+r_D(x_1-x_2+x_3-1))\right]\right.\nonumber\\
&&\left.\cdot E_{enf}(t_b)h_{enf}(x_{1},x_{2},x_{3},b_{1},b_{2})\right\},
\end{eqnarray}

(iii) (S-P)(S+P) operators:
\begin{eqnarray}
\mathcal{M}_{enf}^{SP}&=&-\frac{32}{3}\pi C_{F}m_{B_c}^{4}\int_{0}^{1}d[x]\int_{0}^{1/\Lambda}b_1db_1b_2db_2\phi_{B_c}(x_1,b_1)\phi_{T}(x_{2})\phi_{D}(x_3,b_{1})\nonumber\\
&& \times \left[(r_{D}(x_3-1)-x_1-x_2-x_3+2)E_{enf}(t_{a})h_{enf}(x_{1},(1-x_{2}),x_{3},b_{1},b_{2})\right.\nonumber\\
&&\left.+(r_D(1-x_3)+x_1-x_2)E_{enf}(t_b)h_{enf}(x_{1},x_{2},x_{3},b_{1},b_{2})\right],
\end{eqnarray}
where $C_{F}=4/3$ is the group factor of $SU(3)_c$. The hard scale $t_{a(b)}$ and the functions $E_{enf}$ and  $h_{enf}$ can be found in Appendix A.

Fig. 1(c) and 1(d) are the factorizable annihilation diagrams, whose contributions are

(i) (V-A)(V-A) operators:
\begin{eqnarray}
\mathcal{M}_{af}^{LL}&=&8\sqrt{\frac{2}{3}}C_F\pi f_{B_c}m_{B_c}^{4}\int_{0}^{1}dx_2dx_3\int_{0}^{1/\Lambda}b_2db_2b_3db_3\,\phi_{D}(x_3,b_3)\nonumber\\
&&\times\left\{\left[2\phi_{T}^{s}(x_2)r_{D}r_{T}(x_3+1)+\phi_{T}(x_2)x_3\right]E_{af}(t_c)h_{af1}(x_2,x_3,b_2,b_3)\right.\nonumber\\
&&+\left.\left[\phi_{T}(x_2)(2r_cr_D-x_2)+r_T(-\phi_{T}^{t}(x_2)(2r_D(x_2-1)+r_c)\right.\right.\nonumber\\
&&\left.\left.+\phi_{T}^{s}(x_2)(-2(x_2+1)r_D+r_c))\right]E_{af}(t_d)h_{af2}(x_2,x_3,b_2,b_3)\right\},
\end{eqnarray}

(ii)(S-P)(S+P) operators:
\begin{eqnarray}
\mathcal{M}_{af}^{SP}&=&-16\sqrt{\frac{2}{3}}C_{F}f_{B_c}\pi m_{B_c}^{4}\int_{0}^{1}dx_2dx_3\int_{0}^{1/\Lambda}\phi_{D}(x_3,b_3)\nonumber\\
&&\times\left[(2\phi_{T}^{s}(x_2)r_T+r_D\phi_{T}(x_2)x_3)E_{af}(t_c)h_{af1}(x_2,x_3,b_2,b_3)\right.\nonumber\\
&&\left.+(\phi_{T}(x_2)(2r_D-r_c)+r_T(\phi_{T}^{s}(x_2)(x_2-4r_Dr_c)-\phi_{T}^{t}(x_2)x_2))\right.\nonumber\\
&&\left.\cdot E_{af}(t_d)h_{af2}(x_2,x_3,b_2,b_3)\right],
\end{eqnarray}
with $r_c=m_c/m_{B_c}$. $m_c$ is the mass of the $c$ quark. $t_{c(d)}$, $E_{af}$ and $h_{af1(2)}$ are also listed in Appendix A.

The last two diagrams in Fig.1 are the nonfactorizable annihilation diagrams, whose contributions are

(i) (V-A)(V-A) operators:
\begin{eqnarray}
\mathcal{M}_{anf}^{LL}&=&-\frac{32}{3}C_F\pi m_{B_c}^{4}\int_{0}^{1}d[x]\int_{0}^{1/\Lambda}b_1db_1b_2db_2\phi_{B_c}(x_1,b_1)\phi_{D}(x_3,b_2)\nonumber\\
&&\times\left\{\left[\phi_T(x_2)(1-x_1-x_2-r_b)-r_Tr_D(\phi_T^t(x_2)(x_1+x_2-x_3)\right.\right.\nonumber\\
&&\left.\left.+\phi_T^s(x_2)(x_1+x_2+x_3-2+4r_b))\right]E_{anf}(t_e)h_{anf1}(x_1,x_2,x_3,b_1,b_2)\right.\nonumber\\
&&+\left.\left[\phi_{T}^{s}(x_2)r_Dr_T(-x_1+x_2+x_3+4r_c)+\phi_{T}^{t}(x_2)r_Dr_T(x_1-x_2+x_3)\right.\right.\nonumber\\
&&\left.\left.+\phi_{T}(x_2)(x_3+r_c)\right]E_{anf}(t_f)h_{anf2}(x_1,x_2,x_3,b_1,b_2)\right\},
\end{eqnarray}

(ii) (V-A)(V+A) operators:
\begin{eqnarray}
\mathcal{M}_{anf}^{LR}&=&-\frac{32}{3}C_F\pi m_{B_c}^{4}\int_{0}^{1}d[x]\int_{0}^{1/\Lambda}b_1db_1b_2db_2\phi_{B_c}(x_1,b_1)\phi_{D}(x_3,b_2)\nonumber\\
&&\times\left\{\left[-(\phi_{T}^{t}(x_{2})+\phi_{T}^{s}(x_2))r_{T}(x_{1}+x_{2}-1-r_b)+\phi_{T}(x_{2})r_{D}(x_{3}-1-r_b)\right]\right.\nonumber\\
&&\left.\cdot E_{anf}(t_e)h_{anf1}(x_1,x_2,x_3,b_1,b_2)\right.\nonumber\\
&&\left.+\left[-(\phi_{T}^{s}(x_{2})+\phi_{T}^{t}(x_2))r_{T}(x_{1}-x_{2}+r_{c})-\phi_{T}(x_{2})r_{D}(x_{3}-r_{c})\right]\right.\nonumber\\
&&\left.\cdot E_{anf}(t_f)h_{anf2}(x_1,x_2,x_3,b_1,b_2)\right\},
\end{eqnarray}
with $r_b=m_b/m_{B_c}$. $t_{e(f)}$, $E_{anf}$ and $h_{anf1(2)}$ are also listed in Appendix A.

With the factorization formulae obtained in the above, for these $B_{c}\rightarrow DT$ decays, the total amplitudes containing the Wilson coefficients and CKM elements can be written as
\begin{eqnarray}
\mathcal{A}(B_{c}\rightarrow a_{2}^{+} D^{0})=&&\frac{G_{F}}{\sqrt{2}}\left\{V_{ub}^{*}V_{ud}\mathcal{M}_{enf}^{LL}C_{1}+V_{cb}^{*}V_{cd}(\mathcal{M}_{af}^{LL}a_{1}+\mathcal{M}_{anf}^{LL}C_{1})\right.\nonumber\\
&&\left.-V_{tb}^{*}V_{td}[\mathcal{M}_{enf}^{LL}(C_{3}+C_{9})+\mathcal{M}_{enf}^{LR}(C_{5}+C_{7})+\mathcal{M}_{af}^{LL}(a_{4}+a_{10})\right.\nonumber\\
&&\left.+\mathcal{M}_{af}^{SP}(a_6+a_8)+\mathcal{M}_{anf}^{LL}(C_3+C_9)+\mathcal{M}_{anf}^{LR}(C_5+C_7)]\right\},
\label{1}
\end{eqnarray}
\begin{eqnarray}
\mathcal{A}(B_c\rightarrow K_{2}^{*+}D^{0})=\mathcal{A}(B_{c}\rightarrow a_{2}^{+} D^{0})\mid _{V_{ud}\rightarrow V_{us},V_{cd}\rightarrow V_{cs}, V_{td}\rightarrow V_{ts}, a_{2}^{+}\rightarrow K_{2}^{*+}},
\end{eqnarray}
\begin{eqnarray}
\mathcal{A}(B_c\rightarrow a_{2}^{0} D^{+})=&&\frac{G_{F}}{\sqrt{2}}\frac{1}{\sqrt{2}}\left\{V_{ub}^{*}V_{ud}\mathcal{M}_{enf}^{LL}C_2-V_{cb}^{*}V_{cd}(\mathcal{M}_{af}^{LL}a_{1}+\mathcal{M}_{anf}^{LL}C_{1})\right.\nonumber\\
&&\left.-V_{tb}^{*}V_{td}[\mathcal{M}_{enf}^{LL}(-C_{3}+3a_{10}/2)+\mathcal{M}_{enf}^{LR}(-C_{5}+C_{7}/2)\right.\nonumber\\
&&\left.+\mathcal{M}_{enf}^{SP}(3C_{8}/2)-\mathcal{M}_{af}^{LL}(a_4+a_{10})-\mathcal{M}_{af}^{SP}(a_6+a_8)\right.\nonumber\\
&&\left.-\mathcal{M}_{anf}^{LL}(C_3+C_9)-\mathcal{M}_{anf}^{LR}(C_5+C_7)]\right\},
\end{eqnarray}
\begin{eqnarray}
\mathcal{A}(B_c\rightarrow K_{2}^{*0}D^{+})&=&\frac{G_{F}}{\sqrt{2}}\left\{V_{cb}^{*}V_{cs}(\mathcal{M}_{af}^{LL}a_{1}+\mathcal{M}_{anf}^{LL}C_{1})\right.\nonumber\\
&&\left.-V_{tb}^{*}V_{ts}[\mathcal{M}_{enf}^{LL}(C_{3}-C_{9}/2)+\mathcal{M}_{enf}^{LR}(C_{5}-C_{7}/2)+\mathcal{M}_{af}^{LL}(a_{4}+a_{10})\right.\nonumber\\
&&\left.+\mathcal{M}_{af}^{SP}(a_6+a_{8})+\mathcal{M}_{anf}^{LL}(C_3+C_9)+\mathcal{M}_{anf}^{LR}(C_5+C_7)]\right\},
\end{eqnarray}
\begin{eqnarray}
\mathcal{A}(B_c\rightarrow f_{2}^{q}D^{+})&=&\frac{G_{F}}{\sqrt{2}}\frac{1}{\sqrt{2}}\left\{V_{ub}^{*}V_{ud}\mathcal{M}_{enf}^{LL}C_{2}+V_{cb}^{*}V_{cd}(\mathcal{M}_{af}^{LL}a_{1}+\mathcal{M}_{anf}^{LL}C_1)\right.\nonumber\\
&&\left.-V_{tb}^{*}V_{td}[\mathcal{M}_{enf}^{LL}(C_3+2C_4-C_9/2+C_{10}/2)+\mathcal{M}_{enf}^{LR}(C_5-C_7/2)\right.\nonumber\\
&&\left.+\mathcal{M}_{enf}^{SP}(2C_6+C_8/2)+\mathcal{M}_{af}^{LL}(a_4+a_{10})+\mathcal{M}_{af}^{SP}(a_6+a_8)\right.\nonumber\\
&&\left.+\mathcal{M}_{anf}^{LL}(C_3+C9)+\mathcal{M}_{anf}^{LR}(C_5+C_7)]\right\},
\end{eqnarray}
\begin{eqnarray}
\mathcal{A}(B_c\rightarrow f_{2}^{s}D^{+})&=&\frac{G_{F}}{\sqrt{2}}\left\{-V_{tb}^{*}V_{td}[\mathcal{M}_{enf}^{LL}(C_{4}-C_{10}/2)+\mathcal{M}_{enf}^{SP}(C_6-C_8/2)]\right\},
\end{eqnarray}
\begin{eqnarray}
\mathcal{A}(B_c\rightarrow a_{2}^{0}D_{s}^{+})&=&\frac{G_{F}}{\sqrt{2}}\frac{1}{\sqrt{2}}\left\{V_{ub}^{*}V_{us}\mathcal{M}_{enf}^{LL}C_2-V_{tb}^{*}V_{ts}[\mathcal{M}_{enf}^{LL}3C_{10}/2+\mathcal{M}_{enf}^{SP}3C_{8}/2]\right\},
\end{eqnarray}
\begin{eqnarray}
\mathcal{A}(B_c\rightarrow \bar{K}_{2}^{*0}D_{s}^{+})&=&\frac{G_{F}}{\sqrt{2}}\left\{V_{cb}^{*}V_{cd}(\mathcal{M}_{af}^{LL}a_1+\mathcal{M}_{anf}^{LL}C_{1})-V_{tb}^{*}V_{td}[\mathcal{M}_{af}^{LL}(a_4+a_{10})\right.\nonumber\\
&&\left.+\mathcal{M}_{af}^{SP}(a_6+a_8)+\mathcal{M}_{anf}^{LL}(C_3+C_9)+\mathcal{M}_{anf}^{LR}(C_5+C_7)]\right\},
\end{eqnarray}
\begin{eqnarray}
\mathcal{A}(B_c\rightarrow f_{2}^{q}D_{s}^{+})&=&\frac{G_{F}}{\sqrt{2}}\frac{1}{\sqrt{2}}\left\{V_{ub}^{*}V_{us}\mathcal{M}_{enf}^{LL}C_{2}-V_{tb}^{*}V_{ts}[\mathcal{M}_{enf}^{LL}(2C_4+C_{10}/2)\right.\nonumber\\
&&\left.+\mathcal{M}_{enf}^{SP}(2C_6+C_8/2)]\right\},
\end{eqnarray}
\begin{eqnarray}
\mathcal{A}(B_c\rightarrow f_{2}^{s}D_{s}^{+})&=&\frac{G_{F}}{\sqrt{2}}\left\{V_{cb}^{*}V_{cs}(\mathcal{M}_{af}^{LL}a_{1}+\mathcal{M}_{anf}^{LL}C_{1})-V_{tb}^{*}V_{ts}[\mathcal{M}_{enf}^{LR}(C_{5}-C_{7}/2)\right.\nonumber\\
&&\left.+\mathcal{M}_{enf}^{LL}(C_3+C_4-C_9/2-C_{10}/2)+\mathcal{M}_{enf}^{SP}(C_6-C_8/2)\right.\nonumber\\
&&\left.+\mathcal{M}_{af}^{LL}(a_4+a_{10})+\mathcal{M}_{af}^{SP}(a_6+a_8)+\mathcal{M}_{anf}^{LL}(C_3+C_9)\right.\nonumber\\
&&\left.+\mathcal{M}_{anf}^{LR}(C_5+C_7)]\right\},
\label{2}
\end{eqnarray}

From Eq.(\ref{ffpmix}), we know that
\begin{eqnarray}
\mathcal {A}(B_{c}\rightarrow D^{(*)}f_{2})\,=\,\mathcal
{A}(B_{c}\rightarrow D^{(*)}f_{2}^{q})\cos\theta+\mathcal
{A}(B_{c}\rightarrow D^{(*)}f_{2}^{s})\sin\theta,\\
\mathcal {A}(B_{c}\rightarrow D^{(*)}f_{2}^{\prime})\,=\,\mathcal
{A}(B_{c}\rightarrow D^{(*)}f_{2}^{q})\sin\theta-\mathcal
{A}(B_{c}\rightarrow D^{(*)}f_{2}^{s})\cos\theta,
\end{eqnarray}
with $\theta=7.8^{\circ}$.

The amplitudes of $B_c\rightarrow D^*T$ decay can be decomposed as
\begin{eqnarray}
\mathcal{A}(\epsilon_{D},\epsilon_{T})=i\mathcal{A}^{N}+i(\epsilon_{D}^{T*}\cdot \epsilon_{T}^{T*})\mathcal{A}^{s}+(\epsilon_{\mu\nu\alpha\beta}n^{\mu}v^{\nu}\epsilon_{D}^{T*\alpha}\epsilon_{T}^{T*\beta})\mathcal{A}^{p},
\end{eqnarray}
where $\mathcal{A}^{N}$ contains the contribution from the longitudinal polarizations, while $\mathcal{A}^{s}$ and $\mathcal{A}^{p}$ represent the transversely polarized contributions. $\epsilon_{D}^{T}$ is the transverse polarization vector of $D^{*}$ meson, and $\epsilon_{T}^{T}$ is the vector used to construct the polarization tensors of tensor meson. For each decay process of $B_c\rightarrow D^{*}T$, the amplitudes $\mathcal{A}^N$, $\mathcal{A}^{s}$ and  $\mathcal{A}^{p}$ have the same structures as eqs.(\ref{1})-(\ref{2}), respectively. The factorization formulae for the longitudinal and transverse polarization for the $B_c\rightarrow D^{*}T$ decays are listed in Appendix B.

\section{NUMERICAL RESULTS AND DISCUSSIONS}

The decay width of a $B_c$ meson at rest decaying into $D$
and $T$ meson is
\begin{eqnarray}
\Gamma(B_{c}\rightarrow
DT)\,=\,\frac{|\overrightarrow{P}|}{8\pi
m_{B_c}^{2}}|\mathcal {A}(B_c\rightarrow DT)|\,^{2},
\end{eqnarray}
where the momentum of the final state particle is given by
\begin{eqnarray}
|\overrightarrow
P|\,=\,\frac{1}{2m_{B_c}}\sqrt{\left[m_{B_c}^{2}-(m_{D}+m_{T})^{2}\right]\left[m_{B_c}^{2}-(m_{D}-m_{T})^{2}\right]} .
\end{eqnarray}
The masses and decay constants of tensor mesons needed in the numerical calculations are summarized in
Table \ref{S}.
Other parameters such as QCD scale (GeV), the mass (GeV) and the
lifetime and decay constant of $B_{c}$ meson are
\begin{eqnarray}
&&\Lambda_{\overline{MS}}^{f=4}=0.25,\;m_{B_{c}}=6.286,\;f_{B_c}=0.489,\nonumber\\
&&\tau_{B_{c}}= 0.46 ps,\; \omega_{B_c}=0.6,\,m_{b}=4.8,\;m_{c}=1.5.\nonumber\\
\end{eqnarray}
For the CKM matrix elements, here we adopt the Wolfenstein
parameterization, and take $A=0.808$,
$\lambda=0.2253$, $\bar{\rho}=0.132$ and $\bar{\eta}=0.341$
\cite{jpg37075021}.

\begin{table}[!htbp]
\centering
 \caption{The masses and decay constants of light tensor mesons \cite{TM,TK,zheng1}}
 \vspace{0.3cm}
\begin{tabular}{c!{\;\;\;\;\;\;}c!{\;\;\;\;\;\;}c}
\hline\hline
 \vspace{0.3cm}
 \multirow{1}{*}{Tensor(mass(MeV))} & \multirow{1}{*}{$f_{T}$(MeV)} &\multirow{1}{*}{$f_{T}^{\perp}$(MeV)} \\
\hline
 \vspace{0.1cm}
\multirow{1}{*}{$f_{2}(1270)$}&\multirow{1}{*}{$102\,\pm\,6$}&\multirow{1}{*}{$117\,\pm\,25$}\\
\vspace{0.1cm}
\multirow{1}{*}{$f_{2}^{\prime}(1525)$}&\multirow{1}{*}{$126\,\pm\,4$}&\multirow{1}{*}{$65\,\pm\,12$}\\
\vspace{0.1cm}
\multirow{1}{*}{$a_{2}(1320)$}&\multirow{1}{*}{$107\,\pm\,6$}&\multirow{1}{*}{$105\,\pm\,21$}\\
\vspace{0.3cm}
\multirow{1}{*}{$K_{2}^{*}(1430)$}&\multirow{1}{*}{$118\,\pm\,5$}&\multirow{1}{*}{$77\,\pm\,14$}\\
 \hline\hline
\end{tabular}\label{S}
\end{table}

Like the $\eta\,-\,\eta^{\prime}$ mixing, the isoscalar tensor
states $f_{2}(1270)$ and $f_{2}^{\prime}(1525)$ also have a mixing
and can be given by
\begin{eqnarray}
&&f_{2}\,=\,f_{2}^{q}\cos\theta\,+\, f_{2}^{s}\sin\theta,\nonumber\\
&&f_{2}^{\prime}\,=\,f_{2}^{q}\sin\theta\,-\,f_{2}^{s}\cos\theta,
\label{ffpmix}
\end{eqnarray}
with $f_{2}^{q}\,=\,\frac{1}{\sqrt{2}}(u\bar{u}\,+\,d\bar{d})$,
$f_{2}^{s}\,=\,s\bar{s}$ and the mixing angle
$\theta\,=\,5.8^{\circ}$ \cite{zheng3}, $7.8^{\circ}$
\cite{jpg27807} or $(9\,\pm\,1)^{\circ}$ \cite{jpg37075021}.

For $B_c\rightarrow D^{*}T$ decays, with three kinds of polarization amplitudes, the decay width can be
written as
\begin{eqnarray}
\Gamma(B_c\rightarrow
D^{*}T)\,=\,\frac{|\overrightarrow{P}|}{8\pi
m_{B}^{2}}(\mid\mathcal{A}^{N}\mid^{2}+2(\mid\mathcal{A}^{s}\mid^{2}+\mid\mathcal{A}^{p}\mid^{2})).
\end{eqnarray}

\begin{table}[!h]
\centering
 \caption{Branching ratios (unit:$10^{-6}$) and direct CP asymmetries (unit:$\%$) of $B_{c}\rightarrow DT$ decays calculated in the PQCD approach .}
 \vspace{0.2cm}
\begin{tabular}[t]{l!{\;\;\;\;}c!{\;\;\;\;}c!{\;\;\;\;}c}
\hline\hline

 \multirow{2}{*}{Decay Modes}& \multirow{2}{*}{Class} & \multirow{2}{*}{Br} &
 \multirow{2}{*}{$A_{CP}^{dir}$}\\
 &&&\\
 \hline
\vspace{0.2cm}
\multirow{1}{*}{$B_{c}\rightarrow D^{0}a_{2}^{+}$}&\multirow{1}{*}{A} &\multirow{1}{*}{$2.17_{-0.71\,-0.17\,-0.18}^{+0.83\,+0.17\,+0.20}$} &\multirow{1}{*}{$6.47_{-1.15\,-1.59\,-0.74}^{+1.35\,+5.33\,+0.00}$}\\
\vspace{0.2cm}
\multirow{1}{*}{$B_{c}\rightarrow D^{0}K_{2}^{*+}$}& \multirow{1}{*}{A}& \multirow{1}{*}{$31.9_{-8.76\,-2.86\,-0.54}^{+10.3\,+2.81\,+0.86}$} & \multirow{1}{*}{$-0.44_{-0.15\,-0.22\,-0.02}^{+0.13\,+0.10\,+0.10}$}\\
\vspace{0.2cm}
\multirow{1}{*}{$B_{c}\rightarrow D^{+}a_{2}^{0}$} & \multirow{1}{*}{A} & \multirow{1}{*}{$1.10_{-0.36\,-0.11\,-0.26}^{+0.42\,+0.09\,-0.23}$} & \multirow{1}{*}{$18.2_{-3.77\,-4.65\,-2.30}^{+4.73\,+10.2\,+0.00}$}\\
\vspace{0.2cm}
\multirow{1}{*}{$B_{c}\rightarrow D^{+}K_{2}^{*0}$} & \multirow{1}{*}{A} & \multirow{1}{*}{$31.6_{-9.69\,-2.13\,-0.63}^{+11.3\,+3.10\,+1.01}$} & \multirow{1}{*}{$0.0$}\\
\vspace{0.2cm}
\multirow{1}{*}{$B_{c}\rightarrow D^{+}f_{2}$} & \multirow{1}{*}{A} & \multirow{1}{*}{$1.51_{-0.48\,-0.09\,-0.16}^{+0.58\,+0.12\,+0.14}$} & \multirow{1}{*}{$-9.71_{-3.97\,-5.21\,-1.59}^{+3.45\,+4.09\,+2.70}$}\\
\vspace{0.2cm}
\multirow{1}{*}{$B_{c}\rightarrow D^{+}f_{2}^{\prime}$} & \multirow{1}{*}{A,P} & \multirow{1}{*}{$0.012_{-0.005\,-0.003\,-0.002}^{+0.006\,+0.004\,+0.001}$} & \multirow{1}{*}{$-47.5_{-20.1\,-4.8\,-9.7}^{+16.9\,+10.2\,+9.7}$}\\
\vspace{0.2cm}
\multirow{1}{*}{$B_c\rightarrow D_{s}^{+}a_{2}^{0}$} & \multirow{1}{*}{C} & \multirow{1}{*}{$0.0047_{-0.0007\,-0.0012\,-0.0004}^{+0.0011\,+0.0016\,+0.0006}$} & \multirow{1}{*}{$-2.04_{-0.37\,-1.29\,-0.28}^{+0.34\,+0.62\,+0.58}$}\\
\vspace{0.2cm}
\multirow{1}{*}{$B_c\rightarrow D_{s}^{+}\bar{K}_{2}^{*0}$} & \multirow{1}{*}{A} & \multirow{1}{*}{$1.90_{-0.59\,-0.22\,-0.07}^{+0.67\,+0.20\,+0.09}$} & \multirow{1}{*}{$-1.00_{-0.82\,-0.50\,-0.03}^{+0.76\,+0.72\,+0.00}$}\\
\vspace{0.2cm}
\multirow{1}{*}{$B_c\rightarrow D_{s}^{+}f_{2}$} & \multirow{1}{*}{A,P} & \multirow{1}{*}{$1.87_{-0.40\,-0.44\,-0.06}^{+0.43\,+0.45\,+0.06}$} & \multirow{1}{*}{$2.53_{-0.48\,-0.72\,-0.51}^{+0.51\,+1.45\,+0.10}$}\\
\vspace{0.4cm}
\multirow{1}{*}{$B_{c}\rightarrow D_{s}^{+}f_{2}^{\prime}$} & \multirow{1}{*}{A} & \multirow{1}{*}{$40.9_{-10.7\,-4.17\,-0.81}^{+11.9\,+4.32\,+1.20}$} & \multirow{1}{*}{$-0.11_{-0.02\,-0.06\,-0.00}^{+0.02\,+0.03\,+0.02}$}\\
 \hline\hline
\end{tabular}\label{s1}
\end{table}

The CP averaging branching ratios and the direct CP asymmetries for the considered decay
modes by using the PQCD approach are summarized in
Tables \ref{s1} and \ref{s2}.
The numerical results obtained from perturbative calculation are sensitive to   many parameters. For the theoretical uncertainties in our calculations, we estimated
three kinds of them:
The first errors are caused by the hadronic
parameters of mesons' wave functions, such as the decay constants and the shape parameters of light tensor meson, charmed meson and the $B_{c}$ meson, which are
given in Sec.~\ref{sec:function} and this section. The second errors are estimated from the uncertainty of $\Lambda_{QCD}\,=\,(0.25\,\pm\,0.05)$ GeV and the choice of the hard scales which vary from $0.8t$ to $1.2t$, which characterize the unknown
next-to-leading order QCD corrections. The third error is from
the uncertainties of the CKM matrix elements. It is easy to see that
the most important theoretical uncertainty is caused by the
non-perturbative hadronic parameters, which can be improved by
experiments.

It is easy to find that there are large theoretical uncertainties in any of the individual decay channel calculations mostly due to the shortage of the Tensor meson property. In order to reduce the effects of the choice of input parameters, we define the ratios of the branching ratios between relevant decay modes:
\begin{eqnarray}
&&\frac{Br(B_{c}\rightarrow D^{(*)0}a_{2}^{+})}{Br(B_{c}\rightarrow D^{(*)+}a_{2}^{0})}\sim 2,\\
&&\frac{Br(B_{c}\rightarrow D^{(*)+}K_{2}^{*0})}{Br(B_{c}\rightarrow D^{(*)0}K_{2}^{*+})} \sim \frac{Br(B_{c}\rightarrow D^{(*)+}a_{2}^{0})}{Br(B_{c}\rightarrow D^{(*)+}f_{2})}\sim1,\\
&&\frac{Br(B_{c}\rightarrow D_{s}^{(*)+}\bar{K}_{2}^{*0})}{Br(B_{c}\rightarrow D_{s}^{(*)+}f_{2}^{\prime})}\sim \left(\frac{f_{K_{2}^{*}}^{T}(f_{K_{2}^{*}})V_{cd}}{f_{f_{2}^{\prime}}^{T}(f_{f_{2}^{\prime}})V_{cs}}\right)^{2}\sim \frac{1}{20},\\
&&\frac{Br(B_{c}\rightarrow D^{+}f_{2})}{Br(B_{c}\rightarrow D^{+}K_{2}^{*0})}\sim\left(\frac{1}{\sqrt{2}}\frac{f_{f_{2}}^{T}V_{cd}}{f_{K_{2}^{*}}^{T}V_{cs}}\right)^{2}\sim\frac{1}{20},\\
&&\frac{Br(B_{c}\rightarrow D^{*+}f_{2})}{Br(B_{c}\rightarrow D^{*+}K_{2}^{*0})}\sim\left(\frac{1}{\sqrt{2}}\frac{f_{f_{2}}V_{cd}}{f_{K_{2}^{*}}V_{cs}}\right)^{2}\sim\frac{1}{40}.
\end{eqnarray}
 It is obvious that any significant deviation from the above relations will be a test  of factorization or signal of new physics.

\begin{table}[!h]
\centering
 \caption{Branching ratios (unit:$10^{-6}$), direct CP asymmetries (unit:$\%$) and the percentage of transverse polarizations $R_{T}$(unit:$\%$) of $B_{c}\rightarrow D^{*}T$ decays calculated in the PQCD approach.}
 \vspace{0.2cm}
\begin{tabular}[t]{l!{\;\;}c!{\;\;}c!{\;\;}c!{\;\;}c!{\;\;}c}
\hline\hline

 \multirow{2}{*}{Decay Modes}& \multirow{2}{*}{Class} & \multirow{2}{*}{ Br}& \multirow{2}{*}{$A_{CP}^{dir}$}& \multirow{2}{*}{$R_{T}$}\\
&&&&\\
 \hline
\vspace{0.18cm}
\multirow{1}{*}{$B_c\rightarrow D^{*0}a_{2}^{+}$}&\multirow{1}{*}{A} &\multirow{1}{*}{$7.34_{-1.75\,-0.49\,-0.12}^{+2.05\,+0.99\,+0.24}$} &\multirow{1}{*}{$5.02_{-0.54\,-1.37\,-0.51}^{+0.54\,+1.34\,+0.07}$}&\multirow{1}{*}{$69.8$}\\
\vspace{0.18cm}
\multirow{1}{*}{$B_c\rightarrow D^{*0}K_{2}^{*}$}& \multirow{1}{*}{A}& \multirow{1}{*}{$151_{-26.5\,-10.5\,-3.00}^{+30.1\,+18.2\,+4.69}$} & \multirow{1}{*}{$-0.15_{-0.02\,-0.08\,-0.06}^{+0.02\,+0.05\,+0.03}$} &\multirow{1}{*}{$82.5$}\\
\vspace{0.18cm}
\multirow{1}{*}{$B_c\rightarrow D^{*+}a_{2}^{0}$} & \multirow{1}{*}{A} & \multirow{1}{*}{$3.75_{-0.88\,-0.23\,-0.02}^{+1.05\,+0.49\,+0.05}$} & \multirow{1}{*}{$7.94_{-1.23\,-3.87\,-1.26}^{+1.25\,+4.07\,+0.34}$} & \multirow{1}{*}{$68.2$}\\
\vspace{0.18cm}
\multirow{1}{*}{$B_c\rightarrow D^{*+}K_{2}^{*0}$} & \multirow{1}{*}{A} & \multirow{1}{*}{$158_{-28.5\,-14.9\,-13.4}^{+30.6\,+16.0\,+0.00}$} & \multirow{1}{*}{$0.0$}& \multirow{1}{*}{$80.3$}\\
\vspace{0.18cm}
\multirow{1}{*}{$B_c\rightarrow D^{*+}f_{2}$} & \multirow{1}{*}{A} & \multirow{1}{*}{$3.38_{-0.90\,-0.22\,-0.26}^{+1.03\,+0.43\,+0.33}$} & \multirow{1}{*}{$-2.47_{-1.11\,-5.11\,-0.00}^{+1.01\,+1.55\,+0.82}$} & \multirow{1}{*}{$69.7$}\\
\vspace{0.18cm}
\multirow{1}{*}{$B_c\rightarrow D^{*+}f_{2}^{\prime}$} & \multirow{1}{*}{A} & \multirow{1}{*}{$0.091_{-0.023\,-0.008\,-0.009}^{+0.025\,+0.011\,+0.009}$} & \multirow{1}{*}{$-5.62_{-1.55\,-6.30\,-0.00}^{+1.40\,+4.63\,+0.29}$} & \multirow{1}{*}{$45.3$}\\
\vspace{0.18cm}
\multirow{1}{*}{$B_c\rightarrow D_{s}^{*+}a_{2}^{0}$} & \multirow{1}{*}{C} & \multirow{1}{*}{$0.0051_{-0.0006\,-0.0015\,-0.0004}^{+0.0008\,+0.0022\,+0.0006}$} & \multirow{1}{*}{$-3.81_{-0.17\,-0.81\,-0.51}^{+0.24\,+0.52\,+1.09}$} & \multirow{1}{*}{$12.7$}\\
\vspace{0.18cm}
\multirow{1}{*}{$B_c\rightarrow D_{s}^{*+}\bar{K}_{2}^{*0}$} & \multirow{1}{*}{A} & \multirow{1}{*}{$8.94_{-1.58\,-0.92\,-0.28}^{+1.70\,+0.79\,+0.45}$} & \multirow{1}{*}{$2.30_{-0.14\,-0.45\,-0.01}^{+0.24\,+0.85\,+0.01}$} & \multirow{1}{*}{$82.0$}\\
\vspace{0.18cm}
\multirow{1}{*}{$B_c\rightarrow D_{s}^{*+}f_{2}$} & \multirow{1}{*}{A} & \multirow{1}{*}{$3.60_{-0.38\,-0.51\,-0.08}^{+0.42\,+0.61\,+0.11}$} & \multirow{1}{*}{$2.09_{-0.16\,-0.41\,-0.40}^{+0.15\,+0.39\,+0.10}$} & \multirow{1}{*}{$98.4$}\\
\vspace{0.36cm}
\multirow{1}{*}{$B_c\rightarrow D_{s}^{*+}f_{2}^{\prime}$} & \multirow{1}{*}{A} & \multirow{1}{*}{$190_{-28.1\,-13.2\,-3.88}^{+30.5\,+19.6\,+6.14}$} & \multirow{1}{*}{$-0.036_{-0.003\,-0.012\,-0.001}^{+0.004\,+0.011\,+0.008}$} & \multirow{1}{*}{$89.5$}\\
 \hline\hline
\end{tabular}\label{s2}
\end{table}

For all considered $B_c\rightarrow D^{(*)}T$ decays, the factorizable emission diagrams do not contribute, because the tensor meson can not be produced through local $(V\pm A)$ and $(S\pm P)$ currents. But these decays can get contributions from nonfactorizable and annihilation diagrams. In fact, most of these decays are dominant by the W annihilation diagrams (A) as classified in the tables. There are only four decay channels, which are dominated by the color suppressed (C) or  penguin (P)  diagrams. As we know, usually the annihilation diagrams are power suppressed comparing with the emission diagrams in PQCD approach. But for these considered decay channels, the contributions from the annihilation type diagrams are enhanced by the large CKM elements $V_{cs(d)}$ and thus play a crucial role in amplitudes.

From Table \ref{s1} and \ref{s2}, one can find that most of the predicted branching ratios are in the order of $10^{-6}$ or even bigger. As stated in ref.\cite{haochu1,haochu2}, the LHC experiment, specifically the LHCb, can produce around $5 \times 10^{10}$ $B_c$ events each year. The $B_c$ decays with a decay rate   at the level of $10^{-6}$ can   be detected with a good precision at LHC experiments \cite{su}. On the basis of our predictions, most of these $B_c\rightarrow D^{(*)}T$ decays can be observed in the   experiments soon. 
On the other hand, since the contributions from penguin operators are so small comparing with the contributions from tree operators, the direct CP asymmetries are all very small except $B_c\rightarrow D^{+}f_{2}^{\prime}$. 
For $B_c\rightarrow D^{+}f_{2}^{\prime}$ decay, the tree contributions from $f_{2}^{q}$ term are suppressed by the mixing angle (see \ref{ffpmix}), to be at the same level with penguin contributions from $f_{2}^{s}$ term. The interference is sizable, thus the direct CP asymmetry is around -50\%.
Unfortunately, this decay channel is not accessible easily by current experiments due to a too small branching ratio.

For $B_c\rightarrow D^{*}T$ decays, we also calculate the percentage of the transverse polarization $R_T$, which can be described as
\begin{eqnarray}
R_{T}=\frac{2(|\mathcal{A}^{s}|^{2}+|\mathcal{A}^{p}|^{2})}{|\mathcal{A}^{N}|^{2}+2(|\mathcal{A}^{s}|^{2}+|\mathcal{A}^{p}|^{2})}.
\end{eqnarray}
 Usually from naive factorization expectation, the longitudinal polarizations dominate the branching ratios of $B$ decays. However, from numerical results shown in Table \ref{s2}, one can see that the transverse polarized contributions  are   about at the same level with the longitudinal polarized contributions. In fact, from eq.(\ref{afn},\ref{afs}), we can find that although the transverse polarized contributions are power suppressed, they are also about at the same level with the longitudinal polarized contributions because the two factorizable annihilation diagrams strongly cancel with each other in the longitudinally polarized case. As a result, for these W annihilation diagrams dominant decays, the percentages of the transverse polarization are around $70\%$ or even bigger. This large percentage can be understood as follows \cite{polarization}:  We know that the ``light
quark-unti-quark" pair created from hard gluon are left-handed or
right-handed with equal opportunity. What is more, the $c$ quark from four quark operator is right-handed. So the $D^{*}$ meson can be
longitudinally polarized or transversely polarized with polarization
$\lambda=-1$. For the tensor meson, the anti-quark from four quark
operator is right-handed, and the quark produced from hard gluon can
be either left-handed or right-handed. So the tensor meson can be
longitudinally polarized or transversely polarized with polarization
$\lambda=-1$, because of the additional contribution from the
orbital angular momentum. So the transverse polarization can become
so large with additional interference from other diagrams. For $B_c\rightarrow D_{s}^{*+}f_{2}$, the longitudinal contributions from color suppressed diagrams and W annihilation diagrams strongly cancel with each other, while the transverse contributions can not cancel because the transverse contributions from color suppressed tree diagrams are too small. As a result, the ratio of transverse polarizations becomes as large as 98.4\%. But for the color suppressed dominant $B_c\rightarrow D_s^{*+}a_{2}^{0}$ decay, according to the power counting rules in the factorization assumption, the longitudinal contributions should be dominant due to the quark helicity analysis \cite{helicity1,helicity2}. The ratio is only around 10\%.

\section{SUMMARY}

In this paper, we investigate $B_{c}\rightarrow
D^{(*)}T$ decays within the framework of perturbative
QCD approach. We estimate and calculate the contributions of
different diagrams in the leading order approximation of
$m_{D}/m_{B_c}$ expansion.  Most of these decays are dominant by the W annihilation diagrams, which are only calculable in the pQCD approach.  After calculation, we find that the branching ratios of many decays are in the order of $10^{-6}$ or even bigger, which can be detected in the ongoing experiments. These samples of $B_c$ decays would provide an opportunity to study properties of $B_c$ meson and learn about the modes of the decays with a tensor meson emitted. Most of the direct CP asymmetries are very small because the penguin contributions are too small comparing with the tree contributions. We also predict large ratios of transverse polarizations  around $70\%$ or even bigger for those W annihilation dominant decays.

$\textbf{Acknowledgment}$

 We are very grateful to Dr. Xin Liu for helpful discussions. This Work is supported by
the National
Science Foundation of China under the Grant No.11075168. This research was supported in part by the Project of Knowledge Innovation Program (PKIP) of Chinese Academy of Sciences, Grant No. KJCX2.YW.W10

\begin{appendix}
\section{Related Hard Functions}

In this appendix, we summarize the functions that appear in the analytic formulas in the Section \ref{jiexi}.
The first two diagrams in Fig. 1 are nonfactorizable emission diagrams, whose hard scales $t_{a(b)}$ can be determined by
\begin{eqnarray}
t_{a}=&&\max\{\sqrt{(x_1-r_D^2)(1-x_3)}\,m_{B_c},\sqrt{\mid(x_3-1)[(1-r_D^2)(1-x_2)-(x_1-r_D^{2})]\mid}\,m_{B_c},\nonumber\\
&&1/b_{1},1/b_{2}\},
\end{eqnarray}
\begin{eqnarray}
t_{b}=&&\max\{\sqrt{(x_1-r_D^2)(1-x_3)}\,m_{B_c},\sqrt{\mid(x_3-1)[(1-r_D^2)x_2-(x_1-r_D^{2})]\mid}\,m_{B_c},\nonumber\\
&&1/b_{1},1/b_{2}\}.
\end{eqnarray}
The evolution factors $E_{enf}(t_{a})$ and $E_{enf}(t_{b})$ in the
analytic formulas (see Section \ref{jiexi}) are given by
\begin{eqnarray}
E_{enf}(t)\,=\,\alpha_{s}(t)\exp[-S_{B_c}(t)-S_{T}(t)-S_{D}(t)]| \,_{b_{1}=b_{3}}.
\end{eqnarray}
The Sudakov exponents are defined as
\begin{eqnarray}
S_{B_c}(t)\,=\,s\left(x_{1}\frac{m_{B_c}}{\sqrt{2}},b_{1}\right)\,+\,\frac{5}{3}\int_{1/b_{1}}^{t}\frac{d\bar{\mu}}{\bar{\mu}}\gamma_{q}(\alpha_{s}(\bar{\mu})),
\end{eqnarray}
\begin{eqnarray}
S_{D}(t)\,=\,s\left(x_{3}\frac{m_{B_c}}{\sqrt{2}},b_{3}\right)
\,+\,2\int_{1/b}^{t}\frac{d\bar{\mu}}{\bar{\mu}}\gamma_{q}(\alpha_{s}(\bar{\mu})),
\end{eqnarray}
\begin{eqnarray}
S_{T}(t)\,=\,s\left(x_{2}\frac{m_{B_c}}{\sqrt{2}},b_{2}\right)\,+\,s\left((1-x_{2})\frac{m_{B_{c}}}{\sqrt{2}},b_{2}\right)
\,+\,2\int_{1/b}^{t}\frac{d\bar{\mu}}{\bar{\mu}}\gamma_{q}(\alpha_{s}(\bar{\mu})),
\end{eqnarray}
where the $s(Q,b)$ can be found in the Appendix A in the
ref.\cite{prd63074009}. The function $h_{enf}$ can be given as
\begin{eqnarray}
h_{enf}(x_{1},x_{2},x_{3},b_{1},b_{2})\,&=&\,\left[\theta(b_{2}-b_{1})K_{0}(D_{0}m_{B_c}b_2)I_{0}(D_{0}m_{B_c}b_1)\right.\nonumber\\
&&\left.+\theta(b_{1}-b_{2})K_{0}(D_{0}m_{B_c}b_1)I_{0}(D_{0}m_{B_c}b_2)\right]\nonumber\\
&&\cdot \left\{\begin{array}{ll}
\frac{i\pi}{2}H_{0}^{(1)}\left(\sqrt{|D^{2}|}m_{B}b_{2}\right),& \;\;D^{2}<0;\\
K_{0}\left(Dm_{B}b_{2}\right),&\;\;D^{2}>0,
\end{array}\right.
\end{eqnarray}
with
\begin{eqnarray}
D_{0}^{2}&=&(1-x_3)(x_1-r_{D}^{2}),\\
D^{2}&=&(x_3-1)[(1-r_{D}^{2})x_2-(x_1-r_{D}^{2})].
\end{eqnarray}

For the rest of diagrams, the related functions are summarized as
follows:
\begin{eqnarray}
&&t_{c}\,=\,\max\{\sqrt{(1-r_D^2)x_3}m_{B_c},1/b_{2},1/b_{3}\},\nonumber\\
&&t_{d}\,=\,\max\{\sqrt{x_{2}x_3(1-r_{D}^{2})}m_{B_c},\sqrt{(1-r_D^2)x_2+r_D^2-r_c^{2}}m_{B_c},1/b_{2},1/b_{3}\},\\
&&E_{af}(t)\,=\,\alpha_{s}(t)\cdot \exp[-S_{T}(t)-S_{D}(t)],
\end{eqnarray}
\begin{eqnarray}
h_{af1}(x_{2},x_{3},b_{2},b_{3})\,&=&\,(\frac{i\pi}{2})^{2}H_{0}^{(1)}\left(\sqrt{x_{2}x_{3}(1-r_{D}^{2})}m_{B_c}b_{2}\right)\nonumber\\
&&\left[\theta(b_{2}-b_{3})H_{0}^{(1)}\left(\sqrt{F_{1}^{2}}m_{B_c}b_{2}\right)J_{0}\left(\sqrt{F_{1}^{2}}m_{B_c}b_{3}\right)\right.\,+\nonumber\\
&&\left.\theta(b_{3}-b_{2})H_{0}^{(1)}\left(\sqrt{F_{1}^{2}}m_{B_c}b_{3}\right)J_{0}\left(\sqrt{F_{1}^{2}}m_{B_c}b_{2}\right)\right]\cdot
S_{t}(x_{3}).
\end{eqnarray}
\begin{eqnarray}
h_{af2}(x_2,x_3,b_2,b_3)=h_{af1}(x_2,x_3,b_2,b_3)|\,_ {b_2\longleftrightarrow b_3, F_{1}^{2}\rightarrow F_{2}^{2}},
\end{eqnarray}
with
\begin{eqnarray}
F_{1}^{2}&=&(1-r_D^2)x_3,\\
F_{2}^{2}&=&(1-r_{D}^{2})x_2+r_{D}^{2}-r_{c}^{2}.
\end{eqnarray}
The $S_{t}(x)$ is the Jet function with the expression as
\cite{prd66094010}
\begin{eqnarray}
S_{t}(x)\,=\,\frac{2^{1+2c}\Gamma(3/2+c)}{\sqrt{\pi}\Gamma(1+c)}[x(1-x)]^{c},
\end{eqnarray}
where $c\,=\,0.3$. For the nonfactorizable diagrams, we omit the
$S_{t}(x)$, because it provides a very small numerical effect to the
amplitude \cite{plb555}.
\begin{eqnarray}
t_{e}\,&=&\,\max\{\sqrt{x_{2}x_{3}(1-r_{D}^{2})}m_{B_c},\sqrt{|r_b^{2}-(1-x_3)(1-x_1-(1-r_D^2)x_2)|}m_{B_c},\nonumber\\
&&1/b_{1},1/b_{2}\},\nonumber\\
t_{f}\,&=&\,\max\{\sqrt{x_{2}x_{3}(1-r_{D}^{2})}m_{B_c},\sqrt{|r_c^{2}+x_3(x_1-(1-r_D^2)x_2)|}m_{B_c},1/b_{1},1/b_{2}\},
\end{eqnarray}
\begin{eqnarray}
E_{anf}\,=\,\alpha_{s}(t)\cdot
\exp[-S_{B}(t)-S_{T}(t)-S_{D}(t)]\mid\,_{b_{2}=b_{3}},
\end{eqnarray}
\begin{eqnarray}
h_{anfj}(x_{1},x_{2},x_{3},b_{1},b_{2})\,&=&\,\frac{i\pi}{2}\left[\theta(b_{1}-b_{2})H_{0}^{(1)}\left(Gm_{B_c}b_{1}\right)J_{0}\left(Gm_{B_c}b_{2}\right)\right.\nonumber\\
&&\left.+\theta(b_{2}-b_{1})H_{0}^{(1)}\left(Gm_{B_c}b_{2}\right)J_{0}\left(Gm_{B_c}b_{1}\right)\right]\nonumber\\
&&\times \left\{\begin{array}{ll}
\frac{i\pi}{2}H_{0}^{(1)}\left(\sqrt{|G_{j}^{2}|}m_{B_c}b_{1}\right),&
G_{j}^{2}<0,\\
K_{0}\left(G_{j}m_{B_c}b_{1}\right),& G_{j}^{2}>0,
\end{array}\right.
\end{eqnarray}
with $j=1,2$.
\begin{eqnarray}
G^{2}&=&x_{2}x_{3}(1-r_{D}^{2}),\\
G_{1}^{2}&=&r_b^2-(1-x_3)(1-x_1-(1-r_D^2)x_2),\\
G_{2}^{2}&=&r_c^2+x_3(x_1-(1-r_D^2)x_2).
\end{eqnarray}

\section{factorization formulae for $B_{c}\rightarrow D^{*}T$}

{For longitudinal polarization, the decay amplitude of various diagrams and various effective operators are}
\begin{eqnarray}
\mathcal{M}_{enf}^{LL(N)}&=&\frac{32}{3}\pi C_{F}m_{B_c}^{4}\int_{0}^{1}d[x]\int_{0}^{1/\Lambda}b_1db_1b_2db_2\phi_{B_c}(x_1,b_1)\phi_{T}(x_{2})\phi_{D}(x_3,b_{1})\nonumber\\
&& \times \left\{\left[r_{D}(1-x_{3})-x_{1}-x_{2}+1\right]E_{enf}(t_{a})h_{enf}(x_{1},(1-x_{2}),x_{3},b_{1},b_{2})\right.\nonumber\\
&&\left.+\left[r_{D}(1-x_{3})+x_{1}-x_{2}+x_{3}-1\right]E_{enf}(t_b)h_{enf}(x_{1},x_{2},x_{3},b_{1},b_{2})\right\},
\end{eqnarray}
\begin{eqnarray}
\mathcal{M}_{enf}^{LR(N)}&=&-\frac{32}{3}\pi C_{F}r_{T}m_{B_c}^{4}\int_{0}^{1}d[x]\int_{0}^{1/\Lambda}b_1db_1b_2db_2\phi_{B_{c}}(x_{1},b_1)\phi_{D}(x_3,b_1)\nonumber\\
&&\times\left\{\left[\phi_{T}^{s}(x_2)((r_D-1)(x_1+x_2)-r_Dx_3+1)\right.\right.\nonumber\\
&&\left.\left.+\phi_{T}^t(x_2)(-x_1-x_2+r_D(x_1+x_2+x_3-2)+1)\right]\right.\nonumber\\
&&\left.\cdot E_{enf}(t_{a})h_{enf}(x_{1},(1-x_{2}),x_{3},b_{1},b_{2})\right.\nonumber\\
&&\left.+\left[-\phi_{T}^s(x_2)(x_2-x_1+r_D(x_1-x_2-x_3+1))\right.\right.\nonumber\\
&&\left.\left.+\phi_{T}^t(x_2)(x_2-x_1+r_D(x_1-x_2+x_3-1))\right]\right.\nonumber\\
&&\left.\cdot E_{enf}(t_b)h_{enf}(x_{1},x_{2},x_{3},b_{1},b_{2})\right\},
\end{eqnarray}
\begin{eqnarray}
\mathcal{M}_{enf}^{SP(N)}&=&\frac{32}{3}\pi C_{F}m_{B_c}^{4}\int_{0}^{1}d[x]\int_{0}^{1/\Lambda}b_1db_1b_2db_2\phi_{B_c}(x_1,b_1)\phi_{T}(x_{2})\phi_{D}(x_3,b_{1})\nonumber\\
&& \times \left[(r_{D}(x_3-1)-x_1-x_2-x_3+2)E_{enf}(t_{a})h_{enf}(x_{1},(1-x_{2}),x_{3},b_{1},b_{2})\right.\nonumber\\
&&\left.+(r_D(x_3-1)+x_1-x_2)E_{enf}(t_b)h_{enf}(x_{1},x_{2},x_{3},b_{1},b_{2})\right],
\end{eqnarray}
\begin{eqnarray}
\mathcal{M}_{af}^{LL(N)}&=&8\sqrt{\frac{2}{3}}C_F\pi f_{B_c}m_{B_c}^{4}\int_{0}^{1}dx_2dx_3\int_{0}^{1/\Lambda}b_2db_2b_3db_3\,\phi_{D}(x_3,b_3)\nonumber\\
&&\times\left\{\left[2\phi_{T}^{s}(x_2)r_{D}r_{T}(1-x_3)-\phi_{T}(x_2)x_3\right]E_{af}(t_c)h_{af1}(x_2,x_3,b_2,b_3)\right.\nonumber\\
&&+\left.\left[\phi_{T}(x_2)x_2+r_Tr_c(\phi_{T}^{s}(x_2)-\phi_T^{t}(x_2))\right]E_{af}(t_d)h_{af2}(x_2,x_3,b_2,b_3)\right\},
\label{afn}
\end{eqnarray}
\begin{eqnarray}
\mathcal{M}_{af}^{SP(N)}&=&-16\sqrt{\frac{2}{3}}C_{F}f_{B_c}m_{B_c}^4\pi\int_{0}^{1}dx_2dx_3\int_{0}^{1/\Lambda}\phi_{D}(x_3,b_3)\nonumber\\
&&\times\left[(2\phi_{T}^{s}(x_2)r_T-r_D\phi_{T}(x_2)x_3)E_{af}(t_c)h_{af1}(x_2,x_3,b_2,b_3)\right.\nonumber\\
&&\left.+((\phi_{T}^{s}(x_2)-\phi_{T}^{t}(x_2))r_Tx_2+\phi_{T}(x_2)r_c)\right.\nonumber\\
&&\left.\cdot E_{af}(t_d)h_{af2}(x_2,x_3,b_2,b_3)\right],
\end{eqnarray}
\begin{eqnarray}
\mathcal{M}_{anf}^{LL(N)}&=&-\frac{32}{3}C_F\pi m_{B_c}^{4}\int_{0}^{1}d[x]\int_{0}^{1/\Lambda}b_1db_1b_2db_2\phi_{B_c}(x_1,b_1)\phi_{D}(x_3,b_2)\nonumber\\
&&\times\left\{\left[\phi_T(x_2)(x_1+x_2-1+r_b)+r_Tr_D(\phi_T^t(x_2)(x_1+x_2+x_3-2)\right.\right.\nonumber\\
&&\left.\left.+\phi_T^s(x_2)(x_1+x_2-x_3))\right]E_{anf}(t_e)h_{anf1}(x_1,x_2,x_3,b_1,b_2)\right.\nonumber\\
&&+\left.\left[-\phi_{T}^{s}(x_2)r_Dr_T(x_1-x_2+x_3)+\phi_{T}^{t}(x_2)r_Dr_T(x_1-x_2-x_3)\right.\right.\nonumber\\
&&\left.\left.-\phi_{T}(x_2)(x_3+r_c)\right]E_{anf}(t_f)h_{anf2}(x_1,x_2,x_3,b_1,b_2)\right\},
\end{eqnarray}
\begin{eqnarray}
\mathcal{M}_{anf}^{LR(N)}&=&-\frac{32}{3}C_F\pi m_{B_c}^{4}\int_{0}^{1}d[x]\int_{0}^{1/\Lambda}b_1db_1b_2db_2\phi_{B_c}(x_1,b_1)\phi_{D}(x_3,b_2)\nonumber\\
&&\times\left\{\left[-(\phi_{T}^{t}(x_{2})+\phi_{T}^{s}(x_2))r_{T}(x_{1}+x_{2}-1-r_b)+\phi_{T}(x_{2})r_{D}(x_{3}-1-r_b)\right]\right.\nonumber\\
&&\left.\cdot E_{anf}(t_e)h_{anf1}(x_1,x_2,x_3,b_1,b_2)\right.\nonumber\\
&&\left.+\left[-(\phi_{T}^{s}(x_{2})+\phi_{T}^{t}(x_2))r_{T}(x_{1}-x_{2}+r_{c})-\phi_{T}(x_{2})r_{D}(x_{3}-r_{c})\right]\right.\nonumber\\
&&\left.\cdot E_{anf}(t_f)h_{anf2}(x_1,x_2,x_3,b_1,b_2)\right\},
\end{eqnarray}

For {transverse polarization}, the corresponding decay amplitudes are
\begin{eqnarray}
\mathcal{M}_{enf}^{LL(s)}&=&-\frac{16}{\sqrt{3}}\pi C_{F}m_{B_c}^{4}r_{T}\int_{0}^{1}d[x]\int_{0}^{1/\Lambda}b_1db_1b_2db_2\phi_{B_c}(x_1,b_1)\phi_{D}^{T}(x_3,b_{1})\nonumber\\
&& \times \left\{\left[(\phi_{T}^{a}(x_2)+\phi_{T}^{v}(x_2))(x_1+x_2-1)\right]E_{enf}(t_{a})h_{enf}(x_{1},(1-x_{2}),x_{3},b_{1},b_{2})\right.\nonumber\\
&&\left.+\left[\phi_{T}^{a}(x_2)(x_1-x_2)+\phi_{T}^{v}(x_2)(-2(x_1-x_2+x_3-1)r_D+x_1-x_2)\right]\right.\nonumber\\
&&\left.\cdot E_{enf}(t_b)h_{enf}(x_{1},x_{2},x_{3},b_{1},b_{2})\right\},
\end{eqnarray}
\begin{eqnarray}
\mathcal{M}_{enf}^{LL(p)}=\mathcal{M}_{enf}^{LL(s)}\mid_{\phi_{T}^{a}\leftrightarrow \phi_{T}^{v}},
\end{eqnarray}
\begin{eqnarray}
\mathcal{M}_{enf}^{LR(s)}&=&-\frac{16}{\sqrt{3}}\pi C_{F}m_{B_c}^{4}\int_{0}^{1}d[x]\int_{0}^{1/\Lambda}b_1db_1b_2db_2\phi_{B_c}(x_1,b_1)\phi_{D}^{T}(x_3,b_{1})\phi_{T}^{T}(x_2)\nonumber\\
&& \times \left\{\left[r_{D}(r_{D}-1)(x_3-1)\right]E_{enf}(t_{a})h_{enf}(x_{1},(1-x_{2}),x_{3},b_{1},b_{2})\right.\nonumber\\
&&\left.+\left[r_{D}(r_{D}-1)(x_3-1)\right]E_{enf}(t_b)h_{enf}(x_{1},x_{2},x_{3},b_{1},b_{2})\right\},
\end{eqnarray}
\begin{eqnarray}
\mathcal{M}_{enf}^{LR(p)}=\mathcal{M}_{enf}^{LR(s)},
\end{eqnarray}
\begin{eqnarray}
\mathcal{M}_{enf}^{SP(s)}&=&-\frac{16}{\sqrt{3}}\pi C_{F}m_{B_c}^{4}r_{T}\int_{0}^{1}d[x]\int_{0}^{1/\Lambda}b_1db_1b_2db_2\phi_{B_c}(x_1,b_1)\phi_{D}^{T}(x_3,b_{1})\nonumber\\
&& \times \left\{\left[\phi_{T}^{v}(x_2)(2r_D(x_1+x_2+x_3-2)-x_1-x_2+1)\right.\right.\nonumber\\
&&\left.\left.+\phi_{T}^{a}(x_2)(x_1+x_2-1)\right]E_{enf}(t_{a})h_{enf}(x_{1},(1-x_{2}),x_{3},b_{1},b_{2})\right.\nonumber\\
&&\left.+\left[(\phi_{T}^{a}(x_2)-\phi_{T}^{v}(x_2))(x_1-x_2)\right]E_{enf}(t_b)h_{enf}(x_{1},x_{2},x_{3},b_{1},b_{2})\right\},
\end{eqnarray}
\begin{eqnarray}
\mathcal{M}_{enf}^{SP(p)}=-\mathcal{M}_{enf}^{SP(s)}\mid_{\phi_{T}^{a}\leftrightarrow \phi_{T}^{v}},
\end{eqnarray}
\begin{eqnarray}
\mathcal{M}_{af}^{LL(s)}&=&4\sqrt{2}C_F\pi f_{B_c}r_{D}m_{B_c}^{4}\int_{0}^{1}dx_2dx_3\int_{0}^{1/\Lambda}b_2db_2b_3db_3\,\phi_{D}^{T}(x_3,b_3)\nonumber\\
&&\times\left\{\left[-r_{T}(\phi_{T}^{a}(x_{2})(1-x_3)+\phi_{T}^{v}(x_2)(1+x_3))\right]E_{af}(t_c)h_{af1}(x_2,x_3,b_2,b_3)\right.\nonumber\\
&&+\left.\left[r_{T}(\phi_T^a(x_2)(x_2-1)+\phi_{T}^{v}(x_2)(x_2+1))\right.\right.\nonumber\\
&&\left.\left.-\phi_{T}^{T}(x_2)r_c\right]E_{af}(t_d)h_{af2}(x_2,x_3,b_2,b_3)\right\},
\label{afs}
\end{eqnarray}
\begin{eqnarray}
\mathcal{M}_{af}^{LL(p)}=\mathcal{M}_{af}^{LL(s)}\mid_{\phi_{T}^{a}\leftrightarrow \phi_{T}^{v}},
\end{eqnarray}
\begin{eqnarray}
\mathcal{M}_{af}^{SP(s)}&=&8\sqrt{2}C_F\pi f_{B_c}m_{B_c}^{4}\int_{0}^{1}dx_2dx_3\int_{0}^{1/\Lambda}b_2db_2b_3db_3\,\phi_{D}^{T}(x_3,b_3)\nonumber\\
&&\times\left\{\left[r_{T}(\phi_{T}^{a}(x_2)+\phi_{T}^{v}(x_2))\right]E_{af}(t_c)h_{af1}(x_2,x_3,b_2,b_3)\right.\nonumber\\
&&-\left.\left[r_{D}(\phi_{T}^{T}(x_{2})(r_{D}^{2}-1)+2\phi_{T}^{v}(x_2)r_{T}r_{c})\right]E_{af}(t_d)h_{af2}(x_2,x_3,b_2,b_3)\right\},
\end{eqnarray}
\begin{eqnarray}
\mathcal{M}_{af}^{SP(p)}=\mathcal{M}_{af}^{SP(s)}\mid_{\phi_{T}^{a}\leftrightarrow \phi_{T}^{v}},
\end{eqnarray}
\begin{eqnarray}
\mathcal{M}_{anf}^{LL(s)}&=&\frac{16}{\sqrt{3}}C_F\pi r_D m_{B_c}^{4}\int_{0}^{1}d[x]\int_{0}^{1/\Lambda}b_1db_1b_2db_2\phi_{B_c}(x_1,b_1)\phi_{D}^{T}(x_3,b_2)\nonumber\\
&&\times\left\{\left[-\phi_{T}^{T}(x_2)r_{D}(x_3-1)-2\phi_{T}^{v}(x_2)r_{T}r_b\right]E_{anf}(t_e)h_{anf1}(x_1,x_2,x_3,b_1,b_2)\right.\nonumber\\
&&+\left.\left[\phi_{T}^{T}(x_2)r_Dx_3+2\phi_{T}^{v}(x_2)r_Tr_c\right]E_{anf}(t_f)h_{anf2}(x_1,x_2,x_3,b_1,b_2)\right\},
\end{eqnarray}
\begin{eqnarray}
\mathcal{M}_{anf}^{LL(p)}=\mathcal{M}_{anf}^{LL(s)}\mid_{\phi_{T}^{v}\rightarrow \phi_{T}^{a}},
\end{eqnarray}
\begin{eqnarray}
\mathcal{M}_{anf}^{LR(s)}&=&\frac{16}{\sqrt{3}}C_F\pi m_{B_c}^{4}\int_{0}^{1}d[x]\int_{0}^{1/\Lambda}b_1db_1b_2db_2\phi_{B_c}(x_1,b_1)\phi_{D}^{T}(x_3,b_2)\nonumber\\
&&\times\left\{\left[-(\phi_{T}^{a}(x_2)+\phi_{T}^{v}(x_2))r_T(x_1+x_2-1-r_b)\right.\right.\nonumber\\
&&\left.\left.+\phi_{T}^{T}(x_2)r_D(x_3-1-r_b)\right]E_{anf}(t_e)h_{anf1}(x_1,x_2,x_3,b_1,b_2)\right.\nonumber\\
&&-\left.\left[r_T(\phi_{T}^{a}(x_2)+\phi_{T}^{v}(x_2))(x_1-x_2+r_c)\right.\right.\nonumber\\
&&\left.\left.+\phi_{T}^{T}(x_2)r_D(x_3-r_c)\right]E_{anf}(t_f)h_{anf2}(x_1,x_2,x_3,b_1,b_2)\right\},
\end{eqnarray}
\begin{eqnarray}
\mathcal{M}_{anf}^{LR(p)}=\mathcal{M}_{anf}^{LR(s)}.
\end{eqnarray}
\end{appendix}

\end{document}